\documentclass[preprint, showpacs,superscriptaddress, preprintnumbers,amsmath,amssymb]{revtex4}

\usepackage{graphicx}
\usepackage{dcolumn}
\usepackage{bm}

\begin{document}

\preprint{APS/123-QED}

\title{Interparticle attraction in 2D complex plasmas}

\author{Roman Kompaneets}
\email{kompaneets@mpe.mpg.de}
\affiliation{Max-Planck-Institut f\"ur extraterrestrische Physik, Giessenbachstr. 1, 85748 Garching, Germany}

\author{Gregor E. Morfill}
\affiliation{Max-Planck-Institut f\"ur extraterrestrische Physik, Giessenbachstr. 1, 85748 Garching, Germany}
\affiliation{BMSTU Centre for Plasma Science and Technology, Moscow, Russia}

\author{Alexei V. Ivlev}
\affiliation{Max-Planck-Institut f\"ur extraterrestrische Physik, Giessenbachstr. 1, 85748 Garching, Germany}

\date{\today}

\begin{abstract}
Complex (dusty) plasmas allow experimental studies of various physical processes occurring in classical liquids and solids
by directly observing individual microparticles. A major problem is that the interaction between microparticles is generally
not molecular-like. In this Letter, we propose how to achieve a molecular-like interaction potential in laboratory 2D complex
plasmas. We argue that this principal aim can be achieved by using relatively small microparticles and properly adjusting
discharge parameters. If experimentally confirmed, this will make it possible to employ complex plasmas as a model system
with an interaction potential resembling that of conventional liquids.
\end{abstract}

\maketitle

A complex (or dusty) plasma is a plasma that contains charged
microparticles (dust)~\cite{fortov-book,fortov-ivl-khr-phys.rep-2005, ishihara-j.phys.d.appl.phys-2007,
morfill-ivl-rev.mod.phys-2009, shukla-eli-rev.mod.phys-2009, bonitz-hen-blo-rept.prog.phys-2010, schwabe-2014, ivlev-2014, hartmann-2014}. In laboratory complex
plasmas, the particles are typically of a few $\mu$m in diameter and charged primarily by collection of free ions and
electrons from the plasma~\cite{fortov-ivl-khr-phys.rep-2005, ishihara-j.phys.d.appl.phys-2007,
morfill-ivl-rev.mod.phys-2009, shukla-eli-rev.mod.phys-2009, bonitz-hen-blo-rept.prog.phys-2010}. Such systems allow
experimental studies of various physical processes occurring in liquids and solids by directly observing individual
particles. This idea has inspired a great deal of experiments (see, e.g., studies of shock
waves~\cite{nakamura-bai-shu-phys.rev.lett-1999, samsonov-zhd-qui-phys.rev.lett-2004},
solitons~\cite{samsonov-ivl-qui-phys.rev.lett-2002}, crystallization and melting fronts~\cite{rubin-mor-ivl-nat.phys-2006},
and dislocations in crystals~\cite{nosenko-mor-ros-phys.rev.lett-2011}). In contrast to colloidal
suspensions~\cite{anderson-lek-nature-2002, frenkel-2006}, which can be used for similar purposes, complex plasmas are
characterized by weak damping and therefore allow studying various processes on their intrinsic dynamic time scale.

A major problem in the field of complex plasmas is that the interaction between microparticles is generally not molecular-like,
as the pair potential does not exhibit long-range attraction.
This raises questions as to what extent complex plasmas are suitable to study various fundamental processes
occurring in conventional liquids, such as the liquid-vapor phase transition and critical phenomena~\cite{stanley-book}. In
isotropic complex plasmas, which can be experimentally realized under microgravity
conditions~\cite{ivlev-mor-tho-phys.rev.lett-2008, khrapak-klu-hub-phys.rev.e-2012,pustylnik-tho-mor-j.plasma.phys-2012,
beckers-tri-kro-phys.rev.e-2013, khrapak-tho-cha-phys.rev.e-2013}, the interaction potential is believed to be repulsive at
distances of the order of the interparticle separation~\cite{fortov-ivl-khr-phys.rep-2005, morfill-ivl-rev.mod.phys-2009,
lampe-joy-phys.plasmas-2015}. Under laboratory conditions, the interaction potential $\varphi({\bf r})$ is generally substantially anisotropic
and also non-reciprocal [i.e., $\varphi({\bf r})\not = \varphi(-{\bf r})$, {\it actio} $\not =$ {\it reactio}] due to the presence of plasma
flow~\cite{fortov-ivl-khr-phys.rep-2005, miloch-tru-pec-phys.rev.e-2008, morfill-ivl-rev.mod.phys-2009}. Often, a two-dimensional (2D) complex plasma is
formed in the plane perpendicular to the flow; in this case, the interactions in the monolayer are reciprocal, but believed
to be repulsive, too (see, e.g., the experiment of Ref.~\cite{konopka-mor-rat-phys.rev.lett-2000}).

In this Letter, we use a theoretical foundation for calculating
the pair interaction potential in the presence of ion flow, developed by us before
~\cite{ivlev-zhd-khr-phys.rev.e-2005, kompaneets-kon-ivl-phys.plasmas-2007,
kompaneets-mor-ivl-phys.plasmas-2009, kompaneets-ivl-vla-phys.rev.e-2012},
to make an easy-to-verify prediction as to how to achieve attraction between particles in 2D complex plasmas.
We argue that this can be done in a ground-based experiment with the most common experimental setup.
No external fields need to be applied (in contrast to Refs.~\cite{ivlev-mor-tho-phys.rev.lett-2008, kompaneets-mor-ivl-phys.plasmas-2009}),
as we predict that the attraction can
be achieved by merely adjusting parameters such as the gas pressure, rf
power, and particle size.
Our theoretical approach is robust and realistic
as it is kinetic and accounts for collisions,
the non-Maxwellian velocity distribution of ions, and the electric field
that drives the ion flow; the potential calculated using this approach has been shown to be in excellent
agreement with direct measurements~\cite{kompaneets-kon-ivl-phys.plasmas-2007}. If our prediction is confirmed, it will make it
possible to use 2D complex plasmas as a model system to study fundamental processes in 2D liquids.

Most laboratory experiments on 2D complex plasmas are performed in an rf GEC or similar plasma device, where charged
microparticles are levitated against gravity by the (time-averaged) electric field of the (pre)sheath near the lower
electrode~\cite{fortov-ivl-khr-phys.rep-2005, morfill-ivl-rev.mod.phys-2009}. This region is characterized by the presence
of strong ion flow (with a substantially non-Maxwellian velocity distribution) driven by the field towards the
electrode~\cite{godyak-1982, riemann-j.phys.d.appl.phys-1991, franklin-2003, hershkowitz}; the field
is induced in the plasma to balance the absorption of ions and electrons on the electrode (see Bohm 
criterion~\cite{godyak-1982, riemann-j.phys.d.appl.phys-1991, franklin-2003, hershkowitz}).
The presence of the ion flow is a key factor determining the plasma
shielding and hence the interactions between microparticles. Thus, to describe the shielding of a particle levitated in the (pre)sheath (not in the plasma bulk), 
it is essential
to employ the kinetic description for ions, incorporating the field driving the flow and an ion-neutral collision operator:
\begin{eqnarray}
{\bf v} \cdot \nabla f
+\frac{e}{m} \left(
{\bf E}_{\rm sh} - \nabla \varphi \right)
\cdot \frac{\partial f}{\partial {\bf v}}
 = {\rm St}[f],
\label{kinetic-equation}
\end{eqnarray}
\begin{equation}
\nabla \cdot {\bf E}_{\rm sh} -\nabla^2 \varphi = \frac{e}{\epsilon_0}
\left( \int f \, d{\bf v} -n_{\rm e} +  Q\delta({\bf r})\right).
\label{poisson}
\end{equation}
Here, $f$ is the ion distribution function, ${\bf E}_{\rm sh}$ is the unperturbed (pre)sheath field, which is generally a
function of the vertical coordinate, $\varphi$ is the potential perturbation due to the charged particle, ${\rm St}[f]$ is
the ion-neutral collision operator, $Q$ is the particle charge, $e$ is the elementary charge (ions are assumed to be singly
ionized), $m$ is the ion mass, and $\varepsilon_0$ is the permittivity of free space. The electron density $n_{\rm e}$ is
assumed to either have the Boltzmann
response, $\delta n_{\rm e}=n_{\rm e} e \varphi/T_{\rm e}$, where $T_{\rm e}$ is the electron temperature,
or, as a particular case, to be unperturbed by the particle at all, which corresponds to the limit of infinitely large $T_{\rm e}$.
Note that we neglect ionization as the latter is expected have little effect on the interparticle interactions,
at least at pressures we will consider~\cite{chaudhuri-ieee-2010}.

The principal assumption of our present approach is the homogeneous plasma approximation, in which all unperturbed (by the
particle) quantities do not depend on the vertical coordinate. This is a common assumption, and we have recently shown it to be quite
accurate to describe the shielding at moderate distances, particularly in the direction perpendicular to the
flow~\cite{kompaneets-ivl-nos-phys.rev.e-2014}. Thus, the steady state in our model is determined simply by the balance of
the electric field and collisions, $(e{\bf E}_{\rm sh}/m)\cdot \partial f/\partial {\bf v}={\rm St}[f]$. To calculate
$\varphi({\bf r})$, we use the linear perturbation approximation~\cite{fortov-ivl-khr-phys.rep-2005, morfill-ivl-rev.mod.phys-2009,
ivlev-zhd-khr-phys.rev.e-2005, kompaneets-kon-ivl-phys.plasmas-2007, kompaneets-ivl-nos-phys.rev.e-2014}, i.e., we linearize Eqs.~(\ref{kinetic-equation}) and (\ref{poisson})
with respect to the perturbations induced by the particle.

{\it Cold-neutral approximation.} Since in many experiments the ion flow velocity at the levitation height of the 2D crystal
significantly exceeds the neutral thermal velocity, we start our analysis with the cold-neutral approximation:
\begin{equation}
{\rm St}[f]=-\frac{vf}{\ell}+ \frac{\delta ({\bf v})}{\ell} \int f({\bf r}, {\bf v}') v' \, d{\bf v}',
\label{cmfp-simplified}
\end{equation}
where $\delta({\bf v})$ is the delta-function. The collision length $\ell$ is assumed to be velocity-independent, which is a quite accurate approximation for
superthermal flow velocities and noble gases (typical for experiments with 2D complex plasmas). The dominant collision
mechanism in this case is charge transfer~\cite{lieberman-book}, which is characterized by a weak (logarithmic) velocity
dependence of its cross section~\cite{lieberman-book, smirnov-book}.

The corresponding form of the potential is derived in Ref.~\cite{kompaneets-kon-ivl-phys.plasmas-2007} and given
by Eq.~(6) of that paper. The inclusion of the electron response results in addition of the term
\begin{equation}
\tau_\ell = e E_{\rm sh} \ell/T_{\rm e}
\end{equation}
to the numerator under the square root in the above equation. The potential essentially depends on two dimensionless
numbers, $\tau_\ell$ and
\begin{equation}
\zeta_\ell=\lambda/\ell,
\end{equation}
where
\begin{equation}
\lambda=\sqrt{\frac{\varepsilon_0 E_{\rm sh} \ell}{ n e}}
\end{equation}
and $n$ is the ion density. Let us analyze this
potential in the plane perpendicular to the flow.

We start with the asymptotic expressions for small and large distances. At small distances, the potential obviously becomes
the Coulomb potential, $\varphi=Q/(4\pi \varepsilon_0 r)$. It is easy to show that for $\tau_\ell =0$, the corresponding
range of distances is $r \lesssim \lambda \zeta_\ell^{1/3}$ for $\zeta_\ell \ll 1$, and $r \lesssim \lambda \zeta_\ell$ for
$\zeta_\ell \gg 1$. At larger distances, the potential exhibits a power-law decay,
\begin{equation}
\varphi=\frac{Q\lambda^2\sqrt{2}}{48 \pi \varepsilon_0 r^3}(60 \zeta_\ell^2-1)+O(r^{-4}).
\label{power-law}
\end{equation}
However, numerical calculations show that this asymptotic behavior is reached only at very large distances (e.g., $r \sim
10^3 \lambda$) and that a finite $\tau_\ell$ changes the power-law decay to an exponential decay. Nevertheless,
Eq.~(\ref{power-law}) is helpful in that it already demonstrates the principal possibility of attraction at low
collisionality (small $\zeta_\ell$).

\begin{figure}
\includegraphics[width=7.5cm]{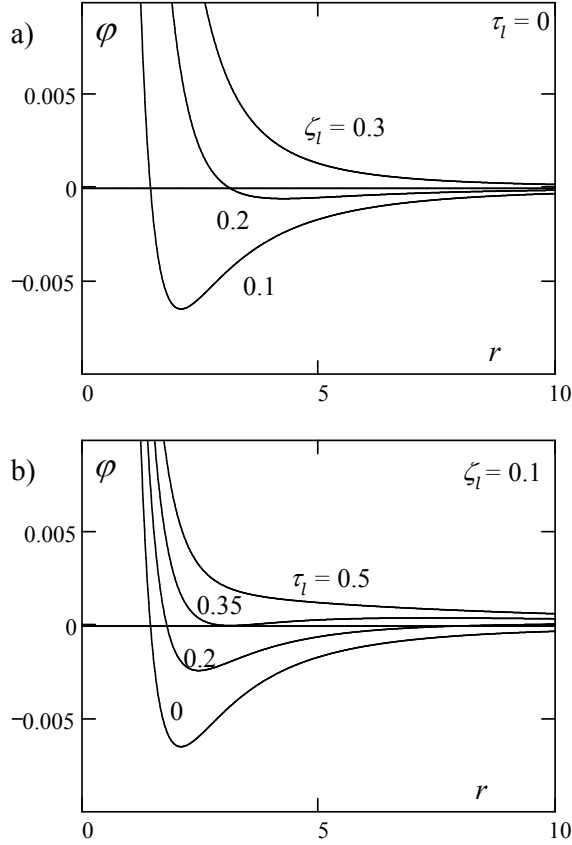}
\caption{Potential in the plane perpendicular to the flow, calculated in the constant-collision-length model under
the cold-neutral approximation. Here, the potential and distance are normalized by $Q/(4 \pi \varepsilon_0 \lambda)$
and $\lambda$, respectively. (a) The limit of absence of the electron response, $\tau_\ell=0$, plotted for various
$\zeta_{\ell}$. (b) Effect of a finite electron response (for $\zeta_\ell=0.1$ and various $\tau_{\ell}$).}
\label{ell-const}
\end{figure}

Figure~\ref{ell-const} shows the results of our numerical analysis of the potential. In the absence of the electron response
($\tau_\ell=0$), the potential is repulsive at small distances and attractive
at large distances for $\zeta_\ell< 0.13$. For $0.13< \zeta_\ell < 0.24$, there is
attraction at intermediate distances and repulsion at small and large distances, while for $\zeta_\ell > 0.24$, the
potential is repulsive at all distances. In the presence of the electron response, our calculations show that the potential
has an attractive well when
\begin{equation}
\zeta_\ell < \frac{0.067}{\tau_\ell+2.8}.
\label{attraction-domain}
\end{equation}
For $\tau_{\ell}\leq2$, this condition is accurate to less than $5$\%.

{\it Role of a finite neutral temperature.} Before we analyze Eq.~(\ref{attraction-domain}) in terms of
experimental parameters such as the gas pressure, plasma density, and particle size, let us first address the role of a finite
neutral temperature. Accurately doing so requires cumbersome velocity calculations, but, to probe into the principal effect,
we simplify the problem by employing the model Bhatnagar-Gross-Krook (BGK) collision operator,
\begin{equation}
{\rm St}[f] = -\nu f +\nu \Phi_{\rm M} \int f({\bf v}') \, d{\bf v}',
\end{equation}
where
\begin{equation}
\Phi_{\rm M}=\frac{1}{(2\pi v_{\rm tn}^2)^{3/2}}\exp\left(-\frac{v^2}{2v_{\rm tn}^2}\right)
\label{neutral-distribution}
\end{equation}
is the normalized Maxwellian velocity distribution of neutrals, $\nu$ is the (velocity-independent) collision frequency, and
$v_{\rm tn}=\sqrt{T_{\rm n}/m}$ is the thermal velocity of neutrals. The corresponding potential is given in
Ref.~\cite{ivlev-zhd-khr-phys.rev.e-2005} and depends on three dimensionless parameters:
\begin{equation}
\tau_\nu=mv_{\rm f}^2/T_{\rm e},
\end{equation}
\begin{equation}
\zeta_\nu=\nu/\omega_{\rm pi},
\end{equation}
and
\begin{equation}
u=v_{\rm f}/v_{\rm tn},
\end{equation}
where $v_{\rm f}=e E_{\rm sh}/(m \nu)$ is the flow velocity. In certain situations (see below), it is convenient to employ
the temperature ratio
\begin{equation}
\tau_{\rm n}=T_{\rm n}/T_{\rm e}
\end{equation}
instead of using $\tau_\nu$.

\begin{figure}
\includegraphics[width=7.5cm]{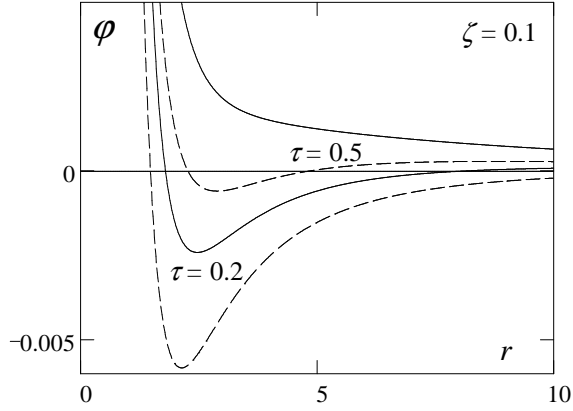}
\caption{Comparison of the constant-collision-length model (solid lines) with the BGK model (dashed lines),
both under the cold-neutral approximation. For the constant-collision-length model ($\zeta=\zeta_\ell$ and $\tau=\tau_\ell$),
the potential and distance are normalized by $Q/(4 \pi \varepsilon_0 \lambda)$
and $\lambda$, respectively. For the BGK model ($\zeta=\zeta_\nu$ and $\tau=\tau_\nu$), the
potential and distance are normalized by $Q \omega_{\rm pi}/(4 \pi \varepsilon_0 v_{\rm f})$
and $v_{\rm f}/\omega_{\rm pi}$, respectively.}
\label{ell-const-vs-bgk}
\end{figure}

In the limit of cold neutrals, $v_{\rm tn} \to 0$ (or $u \to \infty$), the potential (in the plane perpendicular to the
flow) given by the BGK model almost matches the one given by the constant-collision-length model, provided that the
parameters are properly rescaled. Figure~\ref{ell-const-vs-bgk} shows that while the potential curves differ considerably
when $\tau_\ell$ and $\tau_\nu$ are chosen to be the same, changing any of these parameters by about $30$~\% results in
almost perfect matching of the curves. For $\tau_\nu=0$, the potential at large distances is
\begin{equation}
\varphi = \frac{Qv_{\rm f}^2 (\zeta_\nu^2-2)}{4 \pi \varepsilon_0 \omega_{\rm pi}^2 r^3} + O(r^{-4}),
\label{bgk-cold}
\end{equation}
where $\omega_{\rm pi}=\sqrt{n e^2 /(\varepsilon_0 m)}$ is the ion plasma frequency; that is, a similar $r^{-3}$-dependence
is recovered as in the constant-collision-length case.

For finite $\tau_\nu$ (or $\tau_{\rm n}$), the far-field potential at small flow velocities ($u \ll 1$) is
\begin{eqnarray}
\varphi(r )=\frac{Q\tau_{\rm n}\zeta_\nu^2u^2}{8 \pi \varepsilon_0 r(1+\tau_{\rm n})^3}
\nonumber \\
+ \frac{Q\lambda_n^2 u^2(\zeta_\nu^2-2 - 2\tau_{\rm n} \zeta_\nu^2 + 2 \tau_{\rm n}^2)}{4\pi \varepsilon_0
r^3 (1+\tau_{\rm n})^4}
\nonumber \\
+ O \left(\frac{u^4}{r^3} \right)
+O \left( \frac{u^2}{r^4} \right) +O \left[\frac{1}{r}\exp \left( -\frac{r}{\lambda_n} \right) \right],
\label{bgk-long}
\end{eqnarray}
where
\begin{equation}
\lambda_n=\sqrt{\frac{\varepsilon_0 T_n}{n e^2}}
\end{equation}
and $O$ refers to the limit $u \to 0$, $r \to \infty$.
Equation~(\ref{bgk-long}) shows that at very large distances, the potential is always repulsive and Coulomb-like. However,
since in experiments $\tau_{\rm n} \sim 10^{-2}$, the first term is significant only at very large distances; in the second
term, the small parameter $\tau_{\rm n}$ plays little role, so the attraction occurs for $\zeta_\nu <\sqrt{2}$. Note that
for $\tau_\nu = 0$, the potential~(\ref{bgk-long}) reduces exactly to Eq.~(\ref{bgk-cold}), that is, we get the same
far-field potential for small and large flow velocities.

\begin{figure}
\includegraphics[width=7.5cm]{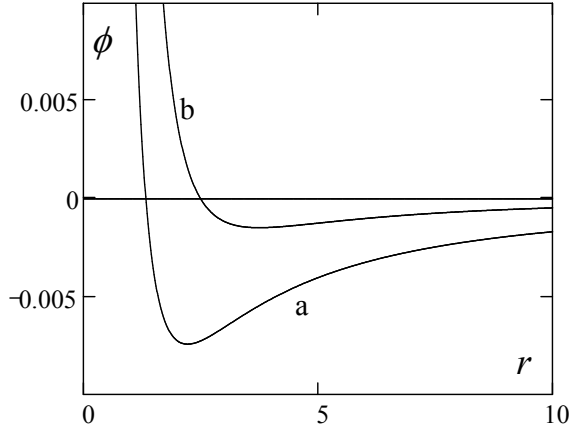}
\caption{Potential for the BGK model and finite realistic values of $u$, $\zeta_\nu$, and $\tau_{\rm n}$.
The potential and distance are normalized as in Fig.~\ref{ell-const-vs-bgk}.
(a) $u=6$, $\zeta_\nu=0.05$, $\tau_{\rm n}=10^{-2}$. (b)
$u=4$, $\zeta_\nu=0.2$, $\tau_{\rm n}=10^{-2}$.}
\label{numerical-bgk}
\end{figure}

In Fig.~\ref{numerical-bgk} we demonstrate that the attraction is present for various realistic combinations of $u$,
$\zeta_\nu$, and $\tau_\nu$.

Thus, since a finite neutral temperature does not suppress the attraction, 
it seems safe to say that all what is needed to achieve the attraction in an experiment is to satisfy the
condition~(\ref{attraction-domain}), provided that the flow velocity substantially exceeds the thermal velocity, i.e., $e
E_0 \ell \gg T_{\rm n}$.

It is noteworthy that in the BGK model, the plasma is unstable with respect to the formation of ion kinetic waves when both
$\zeta_\nu \lesssim 0.3$ and $u \gtrsim 8$; otherwise the plasma is stable~\cite{kompaneets-ivl-vla-phys.rev.e-2012}. Thus,
the attraction emerges before the instability sets in (i.e., before the model itself becomes physically unmeaningful).

We also note that the attraction disappears in the limit $\zeta_\nu \to 0$ as in this case the potential well moves to $r \to
\infty$ and becomes infinitely weak, which indicates that collisions are essential for the attraction. Mathematically, the
limit $\zeta_\nu \to 0$ implies infinitely small ${\bf E}_{\rm sh}$ and $\nu$ but a finite ratio of these (and thus a finite
flow velocity). In this limit, the far-field potential at small flow velocities can be analytically shown to be
\begin{equation}
\varphi(r )=\frac{Q\lambda_n^2 u^2}{ 4 \pi \varepsilon_0r_{}^3} \left[ \frac{2}{(1+\tau_{\rm n})^2}-\frac{\pi}{2(1+\tau_{\rm n})^3}
\right] +\ldots,
\end{equation}
where $\ldots$ are the same $O$-terms as in Eq.~(\ref{bgk-long}). The term is the square brackets in always positive, so
there is always repulsion at large distances. We have numerically verified that the potential remains repulsive at all
distances and finite flow velocities as long as the limit $\zeta_\nu \to 0$ is considered.

{\it Experimental conditions for attraction.} To convert $\zeta_\ell$ and $\tau_\ell$ into experimentally controllable
parameters, we use the vertical force balance, $-QE_{\rm sh}=Mg$ (neglecting the ion drag
force~\cite{fortov-ivl-khr-phys.rep-2005, morfill-ivl-rev.mod.phys-2009}), where $M$ is the particle mass. This yields
\begin{equation}
\zeta_\ell = \sqrt{\frac{\rho a^2 P g \sigma}{3 n z T_{\rm e} T_{\rm n}}}
\label{zeta-exp}
\end{equation}
and
\begin{equation}
\tau_\ell=\frac{\rho a^2ge^2 T_{\rm n}}{3\varepsilon_0 PzT_{\rm e}^2 \sigma},
\label{tau-exp}
\end{equation}
where $a$ is the particle radius, $P$ is the gas pressure, $\rho$ is the particle material density, $z=-Qe/(4\pi \varepsilon_0 a
T_{\rm e})$ is the normalized particle charge, usually of the order of unity~\cite{morfill-ivl-rev.mod.phys-2009}, and $\sigma$ is the
ion-neutral cross-section.

By analyzing the condition for attraction [Eq.~(\ref{attraction-domain})] together with
Eqs.~(\ref{zeta-exp})-(\ref{tau-exp}), we find it rather restrictive, but quite possible to satisfy. For instance, for the
parameters of the experiments of Ref.~\cite{nosenko-ivl-zhd-phys.plasmas-2009}, namely $P=0.66$~Pa, $\rho=1510$~g/cm$^{3}$,
$T_{\rm e}=2.5$~eV, $n \sim 2 \times 10^9$~cm$^{-3}$, and $T_{\rm n}=300$~K, the attraction should occur when the particle
diameter $2a$ is less than about $2.8$~$\mu$m. (This critical size corresponds to $\zeta_\ell \simeq 0.02$ and $\tau_\ell
\simeq 0.5$; to calculate this size, we assumed $z=3$~\cite{fortov-ivl-khr-phys.rep-2005, morfill-ivl-rev.mod.phys-2009} and
$\sigma=6.5\times 10^{-15}$~cm$^{2}$~\cite{kompaneets-kon-ivl-phys.plasmas-2007}.) This is quite a realistic size as the
above experiments were performed with particles of $3.4$--$8.8$~$\mu$m in diameter. Note that the plasma density in the
experiment of Ref.~\cite{nosenko-ivl-zhd-phys.plasmas-2009} was measured in the bulk of the discharge, not at the levitation
height, but this should not affect estimates. Also note that for the above calculated critical size, the flow velocity at
the levitation height is about 6 times the thermal velocity of neutrals, so the cold-neutral approximation should indeed be
applicable. We have also calculated, in the cold-neutral approximation, that the use of 2~$\mu$m particles would result in a
potential well located at about $\simeq 0.13$~mm with the depth $\sim 300 \, T_{\rm n}$, i.e., a deep potential well close
to the particle.

While this Letter focuses on the regime where the flow velocity is much larger than the thermal one, let us note that this
is probably not necessarily required for the attraction. Indeed, in the opposite limit $u \ll 1$, the BGK model still yields
the attraction [see Eq.~(\ref{bgk-long})]. Note that in this regime, the dominant collision mechanism is elastic scattering,
which is not accurately described by the BGK operator, so the attraction is not guaranteed. The attraction condition
$\zeta_\nu <\sqrt{2}$ becomes
\begin{equation}
\sigma P\sqrt{\frac{\varepsilon_0}{2 n_0 e^2 T_{\rm n}}} <1,
\label{attr-low}
\end{equation}
which can be very easily satisfied. For example, for $n=10^9$~cm$^{-3}$ and $T_{\rm}=300$~K, Eq.~(\ref{attr-low}) yields
$P<240$~Pa. However, the condition $u \lesssim 1$ represents a stronger restriction: By using $-QE_{\rm sh}=Mg$, we rewrite
it as
\begin{equation}
\frac{\rho a^2 g e^2}{3 \varepsilon_0 P T_{\rm e} z \sigma} \lesssim 1.
\end{equation}
For $2a=3$~$\mu$m, $T_{\rm e}=2.5$~eV, $\rho=1510$~g/cm$^3$, $z=3$, and $T_{\rm n}=300$~K, this condition yields $P \gtrsim
40$~Pa, which is quite a realistic pressure for experiments with complex plasmas. Larger particles, however, would require a
large pressure to satisfy the condition $u \lesssim 1$.

To conclude, we have theoretically shown that it is possible to obtain a molecular-like interaction potential in 2D complex
plasmas by using relatively small particles and properly adjusting discharge parameters. We hope that our results will
stimulate experimental work in this direction. If experimentally confirmed, the interparticle attraction will make it
possible to employ complex plasmas as a model system with an interaction potential resembling that of conventional liquids.

\begin{acknowledgments}
The work received funding from the European Research Council under the European Union's Seventh Framework Programme, ERC
Grant agreement 267499, and the Russian Scientific Foundation, Project No. 14-43-00053.
\end{acknowledgments}

\end{document}